\documentclass[times,twocolumn,final,authoryear]{elsarticle}

\usepackage{jasr}
\usepackage{framed,multirow}

\usepackage{amssymb}
\usepackage{latexsym}
\usepackage{amssymb}

\usepackage[switch]{lineno}

\usepackage{url}
\usepackage{xcolor}
\definecolor{newcolor}{rgb}{.8,.349,.1}

\usepackage[citebordercolor=white]{hyperref}
\hypersetup{
    colorlinks=true,
    linkcolor=blue
    }

\journal{Advances in Space Research}

\begin{document}

\verso{Given-name Surname \textit{etal}}

\begin{frontmatter}

\title{On the time lag between solar wind dynamic parameters and solar activity UV proxies}

\author[1,2]{R. Reda \corref{cor1}}
\cortext[cor1]{Corresponding author: Raffaele Reda}
\ead{raffaele.reda@roma2.infn.it}
\author[1]{L. Giovannelli}
\ead{luca.giovannelli@roma2.infn.it}
\author[2]{T. Alberti}
\ead{tommaso.alberti@inaf.it}

\address[1]{Department of Physics, University of Rome Tor Vergata, Via della Ricerca Scientifica 1, Rome, 00133, Italy}
\address[2]{INAF - Istituto di Astrofisica e Planetologia Spaziali, Via del Fosso del Cavaliere 100, Rome, 00133, Italy}

\received{}
\finalform{}
\accepted{}
\availableonline{}
\communicated{}

\begin{abstract}
The solar activity displays variability and periodic behaviours over a wide range of timescales, with the presence of a most prominent cycle with a mean length of 11 years. Such variability is transported within the heliosphere by solar wind, radiation and other processes, affecting the properties of the interplanetary medium. The presence of solar activity--related periodicities is well visible in different solar wind and geomagnetic indices, although with time lags with respect to the solar one, leading to hysteresis cycles.
Here, we investigate the time lag behaviour between a physical proxy of the solar activity, the Ca II K index, and two solar wind parameters (speed and dynamic pressure), studying how their pairwise relative lags vary over almost five solar cycles. We find that the lag between Ca II K index and solar wind speed is not constant over the whole time interval investigated, with values ranging from 6 years to $\sim$1 year (average 3.2 years). A similar behaviour is found also for the solar wind dynamic pressure. Then, by using a Lomb-Scargle periodogram analysis we obtain a 10.21-year mean periodicity for the speed and 10.30-year for the dynamic pressure. We speculate that the different periodicities of the solar wind parameters with respect to the solar 11-year cycle may be related to the overall observed temporal evolution of the time lags. Finally, by accounting for them, we obtain empirical relations that link the amplitude of the Ca II K index to the two solar wind parameters.
\end{abstract}


\end{frontmatter}


\section{Introduction}
\label{sec1}
Solar activity, mainly related to changes in the topology and the intensity of the Sun's magnetic field, displays variations on a wide range of timescales, from seconds to millennia \citep[see e.g.,][]{Hathaway2010, Usoskin2017,Vecchio17}. The most prominent variability is represented by the rise and fall of the appearance of magnetic sunspots with a mean periodicity of 11 years, the so-called Schwabe cycle \citep{Schwabe1844}. Such cyclic behaviour is also well visible in all physical proxies of the solar activity, as well as in the synthetic ones. \\
In this regard, the observations of the solar chromosphere in the Ca II K line are of paramount importance \citep[see e.g.][]{Chatzistergos2020}. From an historical point of view they are among one of the longest time series available to characterize the faculae emission linked to the presence of magnetic field in the solar atmosphere, as the first regular observations began in 1904 at Kodaikanal Observatory (India), and later, in 1915, at the Mount Wilson Observatory (USA) \citep{Bertello2016}. On the other hand there is a well established link between the strength of the photospheric magnetic field and the brightness in Ca II K line \citep[see e.g.][]{Schrijver1989, Chatzistergos2019}. The Ca II K index strongly correlate with the magnetic flux, even when no active regions are present in the solar photosphere, whereas the sunspot number correlates only with the fraction of magnetic flux present in sunspots and pores \cite{Bertello2016}. Therefore we exploit in this work the portion of Ca II K dataset, from the \citet{Bertello2016} composite, that overlaps with solar wind direct measurements to assess the level of solar activity from solar UV observations.

The solar activity affects the whole heliosphere since it is carried out from the Sun by the solar wind plasma emitted from the solar corona, by electromagnetic radiation, as well as, by different solar-source mechanisms and structures (as coronal mass ejections, flares, and so on). Such phenomena also affect the Earth, where solar activity induces changes in the near-Earth electromagnetic environment (at short timescales, \citet{Gonzalez1990,Parks04}) and in the overall radiative energy balance (at longer timescales, \citet{North81,Alberti15,Ghil20}).

Because of the direct influence on the terrestrial and circumterrestrial environment, an increased interest is growing in investigating the connection between solar activity and its propagated effects on Earth, also stimulating the birth of new research branches, known as space weather and space climate \citep{Bothmer07}. This also leads an increased attention in connecting the solar activity with solar wind properties, especially with the advent of space missions. Unfortunately, due to the fact that solar wind parameters have been directly measured only since 1964, the relationship between solar activity and solar wind has been initially investigated by means of geomagnetic indices \citep[see e.g.,][]{Hirshberg73, Feynman82}. The presence of a solar-like cycle in the geomagnetic data has been suggested by \citet{Hirshberg73}, who pointed out a cycle not in phase with the 11-year solar activity cycle. Later on many other works investigated the presence of a solar-like cycle in geomagnetic measurements, mostly using the aa-index and its connection with the SunSpot Number (SSN) \citep[see e.g.,][]{Crooker1977, Feynman82, Cliver1996, Echer2004, Richardson2000, Du2011, Richardson2012}.
As a growing number of space missions provided solar wind measurements, the hypothesis of a solar wind cycle, with a characteristic time similar to the 11-year solar cycle, has been directly confirmed on solar wind data. In particular, for the solar wind speed a main periodicity of $\sim$9.6-yr have been found by \citet{ElBorie02, Prabhakaran2002, Dmitriev2009}, while other works reported both shorter (8.3-yr) and longer (10.4-yr) periods \citep[see e.g.,][]{Katsavrias12, Li17}. Similar results have been found also for the solar wind dynamic pressure, with evidence of periodic variability which peaks at 10.2-yr \citep{Dmitriev2009}, 8.3-yr and 11.8-yr \citep{Katsavrias12}. Moreover, periodic variations with a typical time close to the 11-year solar cycle have been observed in other solar wind parameters, such as the helium abundance \citep{Ogilvie1974, Feldman1978, Neugebauer1981, Aellig2001} and the interplanetary magnetic field \citep{Siscoe1978, King1979, Katsavrias12, Dmitriev2009}. \\
Furthermore, a point of great interest in this respect is represented by the observed phase and shape differences between the solar cycle and the one observed in solar wind parameters, as above discussed. Indeed, since the first evidence by \citet{Hirshberg73}, the existence of a time lag has been subsequently confirmed in different papers \citep[see e.g.,][]{Intriligator74, Kohnlein96, Li2016, Venzmer2018, Samsonov2019, Reda2022}, considering different activity proxies. This time-shift could probably be connected to the peak of solar wind High Speed Streams (HSS) from Corotating Interaction Regions (CIRs) during the declining phase of the solar cycle \citep{Tsurutani2006}, when a rise in the number of geomagnetic storms is also observed \citep{Gonzalez1990}.

In particular, in this paper we investigate the relation, on time scales larger than the year, between a physical proxy of the solar activity, the Ca II K index (\citet{Bertello2016} composite), and two solar wind parameters, such as speed and dynamic pressure. Even if the time shift of solar wind parameters with respect to solar activity proxies has been investigated by different authors, this is the first study, to our knowledge, in which the time lag of solar wind parameters to Ca II K index is studied over solar cycle time-scales.  In Sec. \ref{sec2}, we present the dataset used for this analysis and the adopted time window used to filter the short-term variability of the signals. In Sec. \ref{sec3}, we analyze the time lag between the parameters, giving a possible explanation for the observed results. Finally in Sec. \ref{sec4} we discuss the results found.

\section{Dataset and data preparation}
\label{sec2}
\begin{figure*}
    \centering
    \includegraphics[scale=0.45]{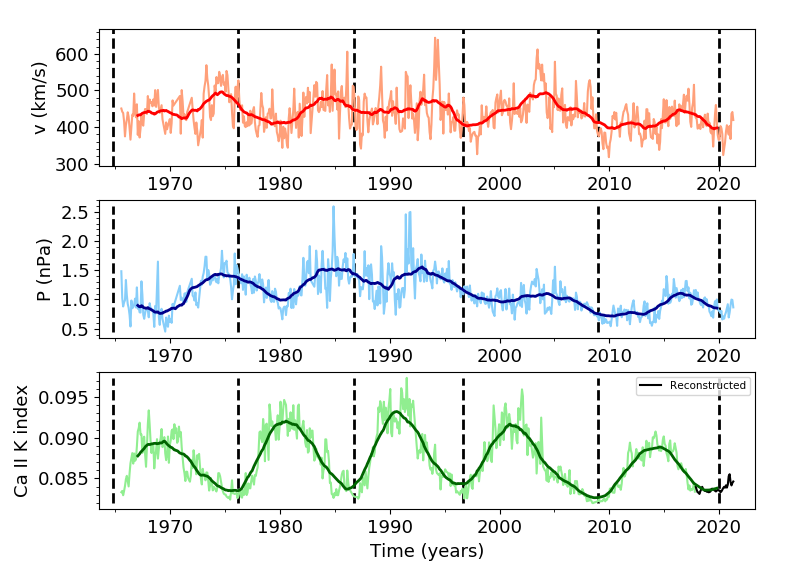}
    \includegraphics[scale=0.45]{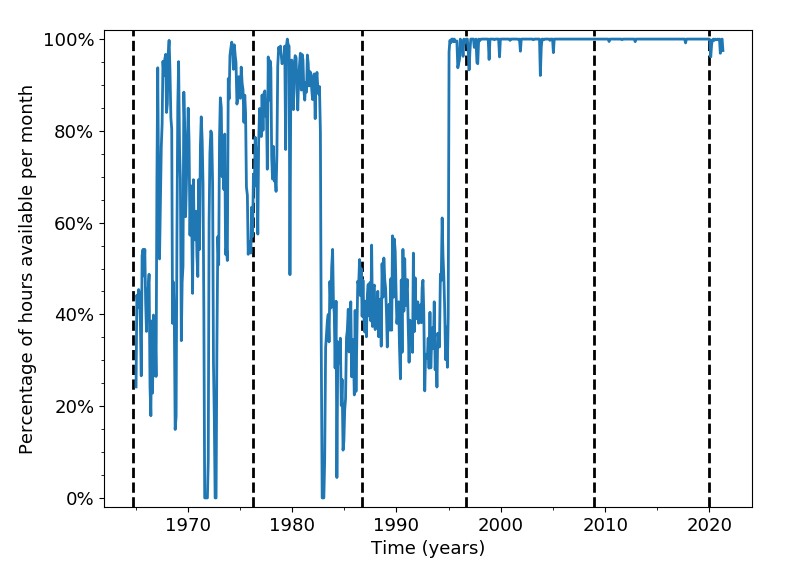}
    \caption{Left panel: Monthly averages of solar wind speed (top), solar wind dynamic pressure (middle) and Ca II K index (bottom). In the bottom panel the black line shows the Ca II K index reconstructed via the Mg II index. The superimposed dark colour lines indicate the 37-month moving averages. The vertical lines are used to separate between the solar cycles.
    Right panel: Solar wind speed coverage, expressed in terms of percentage of hours available per month. Vertical lines are used to separate between the solar cycles.}
    \label{Data}
\end{figure*}

To assess the phase relation between solar activity and solar wind variability, dataset which span over a sufficient long time interval are needed. In this respect, the principal limit is represented by the availability of solar wind measurements. Indeed, direct measurements are available since 1964 within the OMNI database (\url{https://cdaweb.gsfc.nasa.gov/}), which provides to date various near-Earth solar wind parameters with different time resolutions, as collected by different satellite during the time, such as the Interplanetary Monitoring Platform (IMP) (\url{https://ntrs.nasa.gov/citations/19800012928}), the International Sun Earth Explorer (ISEE) \citep{Ogilvie77}, the Advanced Composition Explorer (ACE) \citep{Stone90}, the Wind mission \citep{Lepping95}, and the Geotail one (\url{https://www.isas.jaxa.jp/en/missions/spacecraft/current/geotail.html}) \citep{King2005}. Unfortunately for the first one year and half these data concern only velocity measurements with a lot of time gaps. Thus, we decided to use for this work the data in the time interval July 1965-April 2021, which in any case span over 56 years and cover 5 solar cycles (from 20 to 24). In particular, we start from the hourly-resolution measurements of the ion density $n$ and speed $v$, from which we compute the monthly averages. The right panel of Figure \ref{Data} shows the percentage of available solar wind speed hourly measurements, within the OMNI database, along the selected time interval. The mean percentage of data coverage for each solar cycle (SC) are the following: 62\% for SC~20, 66\% for SC~21, 50\% for SC~22, 100\% for SC~23 and SC~24. Even in the worst coverage periods, we have at least 50\% average coverage.
From the monthly averages of solar wind speed and ion density, we compute another dynamic parameter, the solar wind dynamic pressure, defined as $P=1/2\,m_{p}nv^{2}$, where the proton mass is assumed as the mean ion mass.\\
To quantify the solar magnetic activity several indices have been introduced to date. Among them, we decided to use a physical index of the solar activity, the Ca II K index, which measures the emission of the solar chromosphere \citep{Bertello2016}. In particular the Ca II K index dataset used here is the composite time series described in \citet{Bertello2016}. It contains measurements from the photographic archive of spectroheliograms taken at Kodaikanal Solar Observatory (India, 1907-1987), from the K-line monitor program of disk-integrated measurements from the National Solar Observatory (NSO) at Sacramento Peak (USA, 1988-2006) and finally from the Integrated Sunlight Spectrometer (ISS) on the Synoptic Optical Long-term Investigations of the Sun (SOLIS) at NSO (USA, 2007-2017). While the NSO Sacramento Peak and the SOLIS/ISS datasets are disk-integrated intensity time series, the Kodaikanal one is a plage area series determined by \citet{Tlatov2009} from full-disk Ca II K observations. The three different datasets are combined into a single disk-integrated Ca II K 0.1 nm emission index time series as described in \citet{Bertello2016}. The final Ca II K index composite, available from the National Solar Observatory (NSO) website (\url{https://solis.nso.edu/0/iss/}), contains monthly measurements starting from 1907 and up to October 2017. In order to further extend this dataset to April 2021, as in the case of the solar wind one, other physical indices related to the chromospheric emission can be used. To this scope we make use of the Mg II composite from the University of Bremen (\url{http://www.iup.uni-bremen.de/UVSAT/Datasets/mgii}) as in \cite{Reda2022}.
To obtain smoothed signals and to filter any type of contribution from transient phenomena, related to different time ranges under yearly timescales, we perform a 37-month moving average, following the approach by \citet{Kohnlein96}. The monthly averages of the signals used for this work are shown in the left panel of Figure \ref{Data}, together with their long-term behaviour shown through the superimposed 37-month moving average. For more details regarding the data curation for this work we refer to \cite{Reda2022} where the same dataset and prescriptions were used for a different analysis.

\section{Data analysis}
\label{sec3}
\begin{figure*}
    \centering
    \includegraphics[scale=0.65]{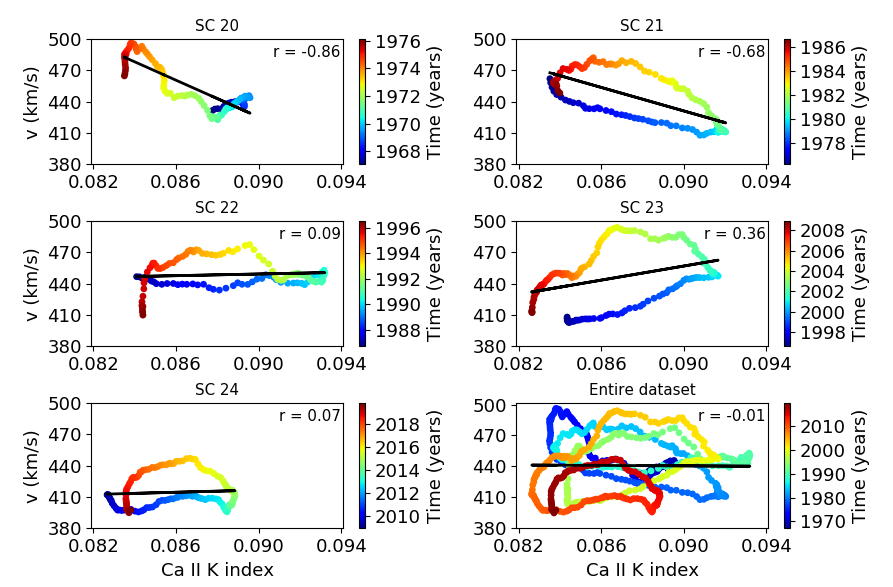}
    \caption{Scatter plots showing the relationship between solar wind speed and Ca II K index for each solar cycle (SC). The scatter plot for the entire period is shown in the bottom-right plot. In each panel the black line represents the linear fit, while the correlation coefficient is indicated on the upper-right.}
    \label{scatter plots v}
\end{figure*}

\begin{figure*}
    \centering
    \includegraphics[scale=0.65]{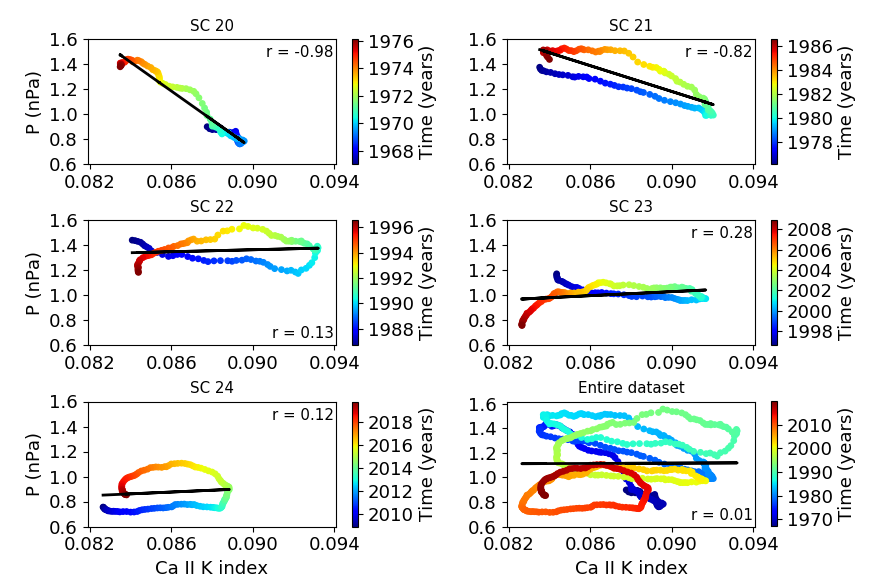}
    \caption{Scatter plots showing the relationship between solar wind dynamic pressure and Ca II K index divided by solar cycles (20-24). The scatter plot for the entire period is shown in the bottom-right plot. In each panel the black line represents the linear fit, while the correlation coefficient is indicated on the upper-right or in the lower-right}
    \label{scatter plots p}
\end{figure*}

As reported by \citet{Li2016} by using the SSN and solar wind speed, there is no significant correlation between solar activity and such solar wind parameter over the whole overlapping time interval of the dataset, which span more than 50-years. We found a similar result for the Ca II K index with both solar wind and dynamic pressure, as shown in the bottom-right panels of Figure \ref{scatter plots v} and \ref{scatter plots p}. However, this result does not prevent to find a relation of solar activity with solar wind speed (or dynamic pressure) over shorter time periods.\\
In order to assess a shorter-time relation, we divide the dataset into solar cycles, according to the official start date (\url{https://wwwbis.sidc.be/silso/cyclesminmax}), and then we compute the correlation coefficient over each of them. The relations over each solar cycle are shown in Figure \ref{scatter plots v} for the solar wind speed and in Figure \ref{scatter plots p} for the solar wind dynamic pressure.
Here, the presence of hysteresis-like phenomena can be noted for both solar wind parameters jointly with the Ca II K index, showing different shapes and widths for each solar cycle. What we can observe by considering the cycle to cycle correlation coefficients is the following:
Ca II K index and solar wind speed are highly anti-correlated over solar cycles 20 and 21, while we observe a passage towards a weak correlation in the subsequent solar cycles 22-23-24. An almost equal time-trend is found also for the dynamic pressure, with the latter that shows a greater absolute value in the correlation coefficient with Ca II proxy, for solar cycles 20 and 21, with respect to that found for the speed. These results highlight a trend of the correlation coefficient over the solar cycles. \\
\begin{figure*}
    \centering
    \includegraphics[scale=0.6]{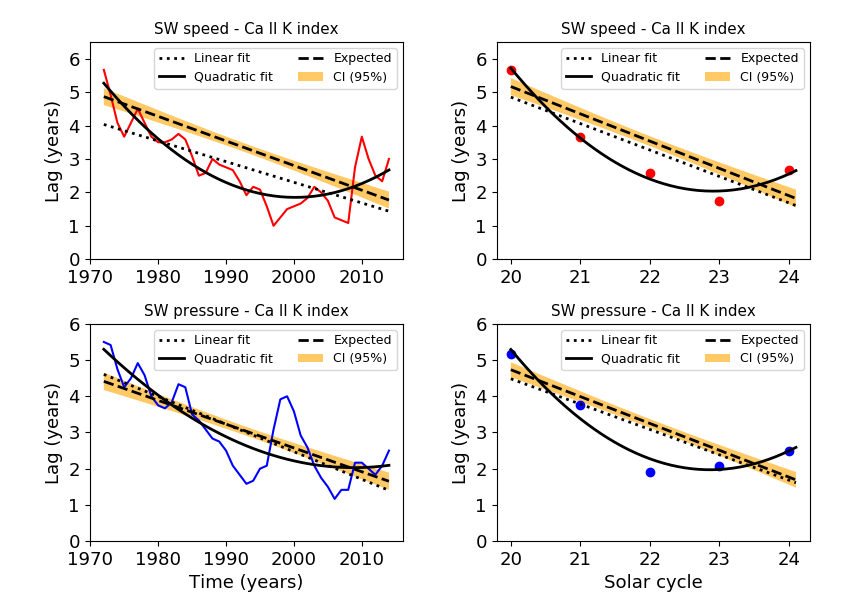}
    \caption{Top: Windowed cross-correlation between Ca II K index and Solar Wind speed over 10-year sliding windows (left panel) and over the different solar cycles (right panel). Bottom: same as above but for the solar wind dynamic pressure. In all the panels the dotted black line shows the linear fit, while the continuous black line represents the quadratic fit. The dashed lines, with their orange shade 95\% confidence internal, show the expected lag for synthetic signals with the same main periodicities of Ca II K index, SW speed and SW dynamic pressure as found by Lomb-Scargle periodogram.}
    \label{windowed cross correlations}
\end{figure*}
In a recent work \citep{Reda2022} we investigated the time lag between Ca II K index and solar wind parameters, considering 37-month averaged data, as in this work. By making use of both cross correlation and mutual information analysis, we found a 3.2-year lag for the solar wind speed with respect to Ca II K index, while a 3.6-year lag is found for the dynamic pressure.
Such analysis has shown the existence of a time lag between Ca II K index and solar wind speed/dynamic pressure, as expected, but the result reported is merely a mean time lag over the whole extension of the dataset. A more continuous information of the lag over the time can be obtained by performing a windowed cross-correlation analysis, as it has been used recently in \cite{Koldobskiy2022} to investigate the time lag between Cosmic-Ray and solar variability. We perform this analysis for two different cases, considering different temporal windows: using a 10-year moving window that it is moved forward by 1 year at each step; over the different solar cycles. In both cases we took the time lag corresponding to the maximum amplitude of each windowed cross-correlation assuming that the solar wind has a delayed response to changes in solar activity, which means that we are considering only positive time lags of the solar wind parameters with respect to Ca II K index. The results of the cross-correlation of Ca II K index with solar wind speed for 10 years sliding windows and over the solar cycles are shown in the top panels of Figure \ref{windowed cross correlations}.
Although the behaviour is characterized by small amplitude peaks, we can see that the solar wind speed's lag decreases almost linearly between 1970 and 1998 and then it starts to grow. A very similar time trend is also found for the cross-correlation over single solar cycles, where we found the maximum lag value of 5.7-year (anti-phase) for the SC 20. Then, we observe a decrease of the lag over the three subsequent cycles, 3.7-year for SC 21, 2.6-year for SC 22 and 1.8-year for SC 23, while it resumes to grow in SC 24 (2.7-year). The analysis carried out between Ca II K index and solar wind dynamic pressure, whose results are shown in the bottom panels of Figure \ref{windowed cross correlations}, shows a quiet similar time trend. In this case, the pressure's lag found through the 10-year sliding windows cross-correlation is characterized by different peaks, which result in the overall quasi-linear decrease over the entire period. When the pressure's lag is seen over the solar cycles, we observe a decrease going from 5.2-year (anti-phase) for SC 20, 3.8-year for SC 21 to 1.9-year for SC 22. A slightly grow of the lag is instead observed for the last two solar cycles, 2.1-year for SC 23 and 2.5-year for SC 24.\\
\begin{figure*}
    \centering
    \includegraphics[scale=0.6]{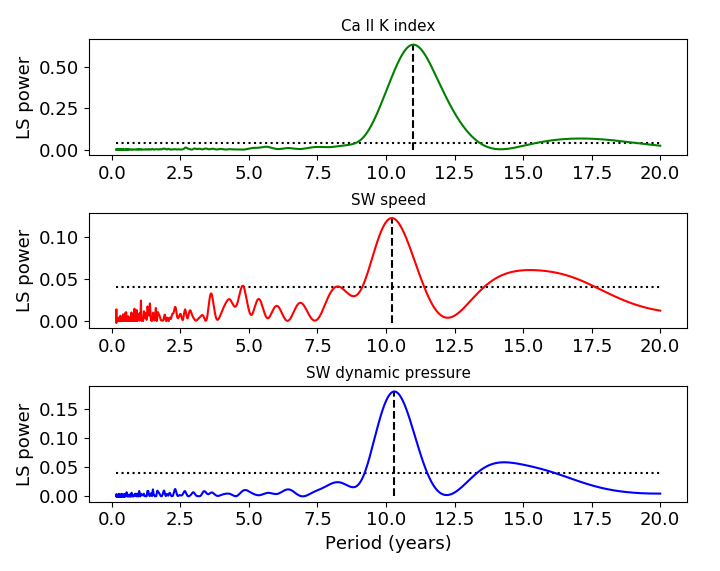}
    \caption{Lomb-Scargle periodograms obtained from monthly values. Top: Ca II K index periodogram showing a main period of 11.00 years. Middle: Periodogram of Solar Wind speed which shows the highest peak at 10.21 years. Bottom: Solar Wind dynamic pressure periodogram showing a main period of 10.30 years. As reference, in each subplots the horizontal dotted line represents the false alarm level for a $1\%$ FAP.}
    \label{periodograms}
\end{figure*}
To investigate the non-constancy and the clear trend found for the lag between Ca II K index and solar wind parameters we compute the Lomb-Scargle periodogram of each time-series for the time interval 1965-2021 starting from their monthly values. We consider the monthly data so that the information on the periodicities takes advantage of a  larger statistic, being the Lomb-Scargle algorithm \citep{Lomb76,Scargle82} optimized to detect periodicity also in unevenly sampled data. The computed Lomb-Scargle periodograms for Ca II K index, solar wind speed and solar wind dynamic pressure are shown is Figure \ref{periodograms}. The analysis reveals a main 11-year period for Ca II K index, which represents the well-known Schwabe cycle of the Sun. Instead, we found for solar wind speed and dynamic pressure a main period of 10.21 and 10.30 years respectively.
The main periods described above are all well-above the false alarm level corresponding to 1\% false alarm probability (FAP), hence they are statistically significant.
We report also other periodicities above the 1\% FAP: a 14.3-yr peak for the solar wind dynamic pressure; 4.8-yr and 15.3-yr peaks for the solar wind speed.\\
These results suggest us the hypothesis that the different main periods found could explain the lag trend previously uncovered in the time-series used for this work. In particular, the lag progression over the solar cycles could be related to the slightly shorter periods of the solar wind parameters with respect to the 11-year associated to the solar UV variations.
To investigate this hypothesis on the different mean periods we build up two synthetic signals having the same periodicities as the solar wind speed and dynamic pressure and we explore the expected lag, with its 95\% confidence interval, between Ca II K index and these two synthetic signals (Figure \ref{windowed cross correlations}).
The two synthetic signals, acting as solar wind speed and dynamic pressure,
are build up as follows: we start from the 37-month averages of Ca II K index and we stretch the time series in order to match the measured main periods of solar wind speed and dynamic pressure (10.21-yr and 10.30-yr respectively); furthermore, we take into account the lag found for solar cycle 20, so that the two synthetic signals have the same initial lag as the original time series. To stretch the time series we performed an interpolation considering as stretching factor the ratio between the main period of solar wind speed (dynamic pressure) and that of Ca II K index. 
The same windowed cross-correlation procedure used on the data is then applied to synthetic time series to obtain the expected lag in this simplified scenario. Our hypothesis seems plausible for the lag of the solar wind dynamic pressure with respect to Ca II K index (bottom panels in Figure \ref{windowed cross correlations}), for which the expected lag is almost always in agreement, within the confidence interval, to the lag's linear fit over time and solar cycles. Looking at the windowed cross-correlation for the solar wind speed (top panels in Figure \ref{windowed cross correlations}), the hypothesis does not work as good as for the dynamic pressure but, especially on solar cycles, the linear fit to the lags does not differ much from the confidence interval of the expected one. 
Therefore, the difference in the main period of Ca II K index and solar wind speed/dynamic pressure time series may be a possible explanation for the phase differences evolution over the time.\\
As a final step we compare our results from the Lomb-Scargle periodograms to other studies on solar wind parameters and geomagnetic indices periodicities.
Focusing the attention only to solar cycle timescales periodicities, a 9.6-year periodicity for solar wind speed for the time interval 1973-2000 has been reported by \citet{ElBorie02}. This period is not very different from what we found (10.21 yrs), but we have to consider that he used only 27 years of data while our dataset is 56 years long. Different values were obtained by \citet{Katsavrias12} computing both a wavelet analysis and Lomb-Scargle periodograms of solar wind parameters for the time interval 1966-2010, founding a period of 8.3-year for solar wind speed and 8.3 and 11.8 years for dynamic pressure. Instead, the main periodicity we report for the solar wind speed is very close to the one found by \citet{Li17} which, by performing a Lomb-Scargle analysis on the daily means for the time interval 1963-2015, got a statistically significant peak at 10.40 years. The main period we found for the solar wind dynamic pressure is quite in agreement with the analysis by \citet{Dmitriev2009}, which found a periodicity peaks at 10.2-year.
Moreover, the peak at 15.3-year that we found for the solar wind speed, is quite similar to that reported for the same solar wind parameter by \citet{Prabhakaran2002} and \citet{Li17}, which found prominent peaks at $\sim$16 years and 15.79 years, respectively.\\
\begin{figure*}[h!]
    \centering
    \includegraphics[scale=0.6]{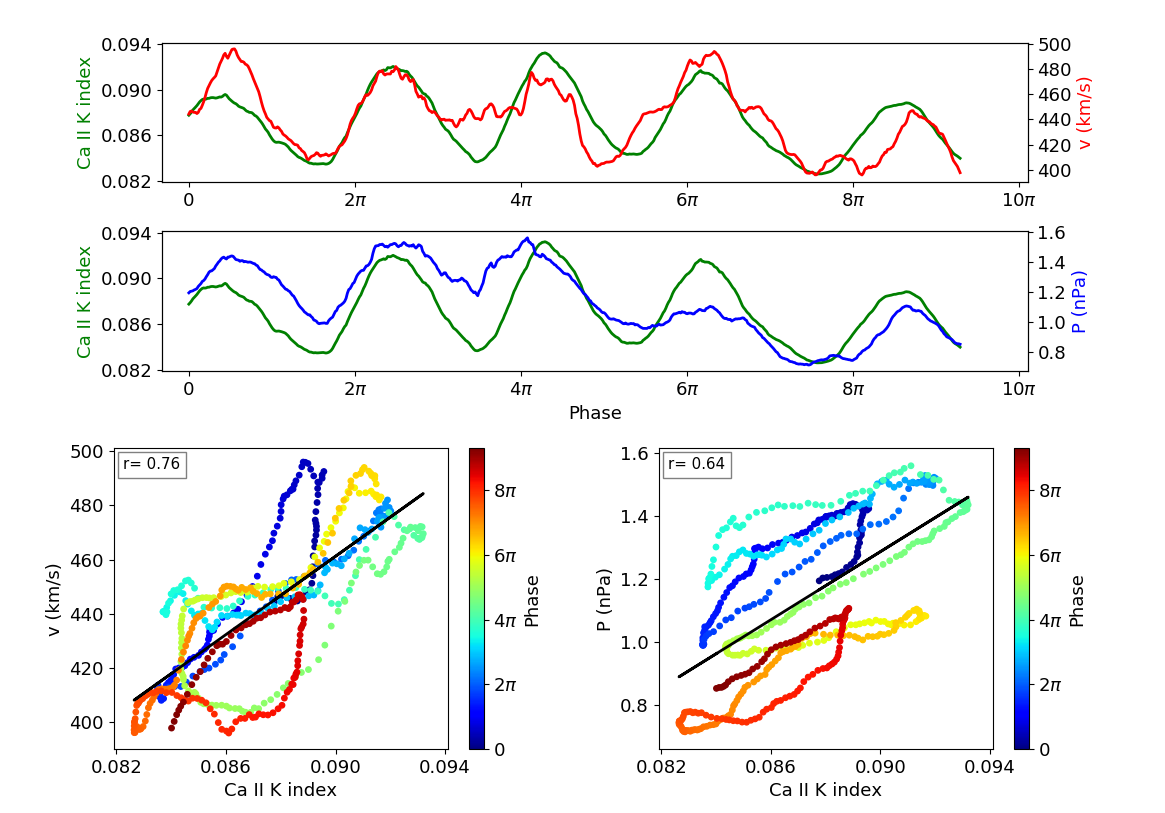}
    \caption{Time warped time series as a function of the phase of the Schwabe 11-year cycle. Top: Solar wind speed (red) and Ca II K index (green) plotted against the phase of the 11-year solar cycle. Middle: same as the top panel but for dynamic pressure (blue). 
    Bottom: Scatter plot relative to time warped time series, showing the relation between Ca II K index and the solar wind speed (left panel) and the solar wind dynamic pressure (right panel), respectively. The correlation coefficients are 0.76 and 0.64. The color map shows the evolution of the relation over the phase of the 11-year cycle.}
    \label{Normalized signals}
\end{figure*}
However, the correlation of Ca II K index with solar wind parameters is influenced by the lags among the signals possibly due to the different main periodicities, which lead to phase asynchrony over the time interval investigated. Thus, in order to find a stronger correlation relation on the signal amplitudes we perform a time warping transformation to detect an optimal match on the time series pairs, a technique already applied to time-series of solar-wind data \citep{Laperre2020,Samara2022}. In this first attempt we perform a uniform time warp over the whole time series taking advantage of the results from the Lomb-Scargle periodograms, leaving the dynamic time warping approach and a more deep phase analysis for a future work.
Here we perform the time warping on the Ca II K index, the SW speed and the dynamic pressure time-series by normalizing the time axis of each quantities to their main periods, as they are found through the Lomb-Scargle analysis (11 years for Ca II K index, 10.21 years for SW speed and 10.30 years for dynamic pressure). 
This way, we introduce three new time series that we plot against the phase of the Schwabe 11-year cycle, shown in top and middle panels of Figure \ref{Normalized signals}.
The amplitude relationship of Ca II K index with solar wind speed and dynamic pressure is shown in the bottom panels of Figure \ref{Normalized signals}, where we use a color map to highlight how the relation changes with the phase of the Schwabe 11-year cycle.
The adopted time warping technique leads to correlated time series, with the correlation coefficient of Ca II K index that increases up to 0.76 with solar wind speed and to 0.64 with dynamic pressure. The empirical linear relations between the two pairs of quantities are the following:
\begin{equation}
\label{eq1}
P\,(\mathrm{nPa}) = (54.0 \pm 2.7)\;\mathrm{Ca\,II\,K} - (3.6 \pm 0.2); 
\end{equation}
\begin{equation}
\label{eq2}
v\,(\mathrm{km/s}) = (7222 \pm 254)\;\mathrm{Ca\,II\,K} - (189 \pm 22). 
\end{equation}
We emphasize that, in the scatter plot of the Ca II K index with the solar wind dynamic pressure (bottom-right panel of Figure \ref{Normalized signals}), the presence of hysteresis patterns is clearly visible. In particular, it is interesting to note that the hysteresis loops do not follow the same path on all solar cycles, which results anticlockwise for SC 20 and 21, clockwise for SC 22 and 23 and again anticlockwise for SC 24. We left a deeper analysis on the origin of this hysteresis behavior in a future detailed work.
\section{Discussion and conclusions}
\label{sec4}
In this work we studied the lag among the time series of the 11-year solar cycle, by means of a physical proxy of the solar activity, the Ca II K index, and the solar-like cycle observed in two solar wind dynamic parameters, i. e. speed and dynamic pressure. 
To perform such analysis we adopted a 37-month moving average to filter the short-term and yearly time-scales variability, maintaining a monthly temporal resolution. Although there is no pairwise correlation in the overall 56-year time interval investigated, a correlation arise if the analysis is performed over the individual solar cycles. Moreover, a statistically significant global correlation  is found if the presence of a time lag is considered. Indeed, we found a strong anti-correlation of Ca II K index with both solar wind parameters over the solar cycles 20 and 21, while we observe a slightly correlation in the subsequent cycles 22, 23 and 24.
A possible theoretical explanation for this behaviour could be found in a feedback mechanism of the solar wind on the solar dynamo. Recently, by using a 2.5-dimensional dynamo-solar wind coupling model, \citet{Perri2021} showed the tendency of solar wind proxies to depart from correlation with solar cycle once such feedback mechanism is turned on. As this model has been tested on a young solar-like star with short dynamo period, further investigation is needed to support this hypothesis. \\
It is worth to point out that the present analysis might be influenced by the partial time coverage of solar wind data before 1996. Furthermore, the different dataset used to obtain the Ca II K composite time series might influence the analysis. Nevertheless, we expect that any significant influence of the aforementioned properties of the database used in this study would show a significant difference performing the same analysis for two distinct sub datasets for the solar cycles 20-21 and 22-23-24. Apart for an expected difference in the retrieved lag for the two subsets the analysis does not show any significant difference with the one here presented for the full dataset. In particular the empirical linear relations are compatible in both subsets with relations \ref{eq1} and \ref{eq2}. \\
In a recent work \citep{Reda2022}, by using both a cross correlation and a mutual information analysis, we investigated the mean time lag among the same signals used for this study, pointing out that solar wind speed and dynamic pressure lag the Ca II K index by 3.2-year and 3.6-year respectively. Here, by adopting a windowed cross correlation analysis, over each solar cycle and over 10-year sliding windows, we found that such time lags are not constant over the time. For both solar wind parameters, the lag decreases from the $\sim$5.5-year of solar cycle 20 down to the $\sim$2-year of solar cycle 23, growing again towards the $\sim$2.5-year of solar cycle 24. We believe that such temporal evolution of the time lag is very interesting and it deserves to be studied in more detail. As a possible explanation for the overall observed time trend of the lag we propose a simple model which take into account for the different main periodicities of the signals, as found with Lomb-Scargle periodograms (11 years for Ca II K index, 10.21 for solar wind speed and 10.30 years for solar wind dynamic pressure).
To do that, we performed a uniform time warping transformation on each time series. By using this approach and by comparing the signals amplitudes along the phase of the 11-year solar cycle, we found a good correlation of the Ca II K index with solar wind speed (r=0.76). A similar result is found also considering the correlation of Ca II K index with solar wind dynamic pressure (r=0.64), whose amplitude relation show the presence of hysteresis loops with clockwise and anticlockwise paths. \\
The presence of hysteresis-like phenomena has been shown in the relation between various solar activity indices by different authors, mainly related to the different paths followed by each indicators in the ascending and descending phases of the solar cycle. By analyzing the hysteresis patterns of the Galactic Cosmic Rays (GCRs) intensity and International SSN, \citet{Kane2003} concluded that odd cycles show broad hysteresis loop, while even ones show narrow loops. An hysteresis behaviour for the Coronal Mass Ejection (CME) speed index has been pointed out when related to different activity indices (e.g. Mg II index, TSI, SSN) \citep{Ozguc2012}, geomagnetic indices, interplanetary magnetic field and solar wind speed \citep{Ozguc2016}, concluding that the hysteresis pattern could be related to the different contribution of magnetic field at different scales along the diverse phases of the solar cycle. Hysteresis phenomena are also reported when GCRs are related to SSN \citep{Ross2019} and CME speed index \citep{Sarp2019}. The investigation of hysteresis relationships between solar and environments near-Earth indices can shed light on the processes taking place along the propagation of solar activity variability within the heliosphere. A deeper investigation of the source mechanisms of the observed hysteresis is left for a forthcoming paper.

\section{Acknowledgments}
R.R. is a PhD student of the PhD course in Astronomy, Astrophysics and Space Science, a joint research program between the University of Rome “Tor Vergata”, the Sapienza University of Rome and the National Institute of Astrophysics (INAF).
The authors thanks Francesco Berrilli, Luca Bertello, Dario Del Moro, Maria Pia Di Mauro, Piermarco Giobbi and Valentina Penza for useful discussions. 

\bibliographystyle{model5-names}
\biboptions{authoryear}
\bibliography{refs}

\begin{thebibliography}{56}
\expandafter\ifx\csname natexlab\endcsname\relax\def\natexlab#1{#1}\fi
\providecommand{\url}[1]{\texttt{#1}}
\providecommand{\href}[2]{#2}
\providecommand{\path}[1]{#1}
\providecommand{\DOIprefix}{doi:}
\providecommand{\ArXivprefix}{arXiv:}
\providecommand{\URLprefix}{URL: }
\providecommand{\Pubmedprefix}{pmid:}
\providecommand{\doi}[1]{\href{http://dx.doi.org/#1}{\path{#1}}}
\providecommand{\Pubmed}[1]{\href{pmid:#1}{\path{#1}}}
\providecommand{\bibinfo}[2]{#2}
\ifx\xfnm\relax \def\xfnm[#1]{\unskip,\space#1}\fi
\bibitem[{Aellig et~al.(2001)Aellig, Lazarus \& Steinberg}]{Aellig2001}
\bibinfo{author}{Aellig, M.~R.}, \bibinfo{author}{Lazarus, A.~J.}, \&
  \bibinfo{author}{Steinberg, J.~T.} (\bibinfo{year}{2001}).
\newblock \bibinfo{title}{The solar wind helium abundance: Variation with wind
  speed and the solar cycle}.
\newblock {\it \bibinfo{journal}{Geophysical Research Letters}\/},  {\it
  \bibinfo{volume}{28}\/}\bibinfo{issue}{(14)}, \bibinfo{pages}{2767--2770}.
  \DOIprefix\doi{https://doi.org/10.1029/2000GL012771}.
\bibitem[{{Alberti} et~al.(2015){Alberti}, {Primavera}, {Vecchio}, {Lepreti} \&
  {Carbone}}]{Alberti15}
\bibinfo{author}{{Alberti}, T.}, \bibinfo{author}{{Primavera}, L.},
  \bibinfo{author}{{Vecchio}, A.}, \bibinfo{author}{{Lepreti}, F.}, \&
  \bibinfo{author}{{Carbone}, V.} (\bibinfo{year}{2015}).
\newblock \bibinfo{title}{{Spatial interactions in a modified Daisyworld model:
  Heat diffusivity and greenhouse effects}}.
\newblock {\it \bibinfo{journal}{Physical Review E}\/},  {\it
  \bibinfo{volume}{92}\/}\bibinfo{issue}{(5)}, \bibinfo{pages}{052717}.
  \DOIprefix\doi{10.1103/PhysRevE.92.052717}.
\bibitem[{{Bertello} et~al.(2016){Bertello}, {Pevtsov}, {Tlatov} \&
  {Singh}}]{Bertello2016}
\bibinfo{author}{{Bertello}, L.}, \bibinfo{author}{{Pevtsov}, A.},
  \bibinfo{author}{{Tlatov}, A.}, \& \bibinfo{author}{{Singh}, J.}
  (\bibinfo{year}{2016}).
\newblock \bibinfo{title}{{Correlation Between Sunspot Number and Ca ii K
  Emission Index}}.
\newblock {\it \bibinfo{journal}{Sol. Phys.}\/},  {\it
  \bibinfo{volume}{291}\/}\bibinfo{issue}{(9-10)}, \bibinfo{pages}{2967--2979}.
  \DOIprefix\doi{10.1007/s11207-016-0927-9}.
  \href{http://arxiv.org/abs/1606.01092}{\tt arXiv:1606.01092}.
\bibitem[{{Bothmer} \& {Daglis}(2007)}]{Bothmer07}
\bibinfo{author}{{Bothmer}, V.}, \& \bibinfo{author}{{Daglis}, I.~A.}
  (\bibinfo{year}{2007}).
\newblock {\it \bibinfo{title}{{Space Weather -- Physics and Effects}}\/}.
\newblock \bibinfo{publisher}{Springer Berlin, Heidelberg}.
\newblock \DOIprefix\doi{10.1007/978-3-540-34578-7}.
\bibitem[{{Chatzistergos} et~al.(2020){Chatzistergos}, {Ermolli}, {Krivova},
  {Solanki}, {Banerjee}, {Barata}, {Belik}, {Gafeira}, {Garcia}, {Hanaoka},
  {Hegde}, {Klime{\v{s}}}, {Korokhin}, {Louren{\c{c}}o}, {Malherbe},
  {Marchenko}, {Peixinho}, {Sakurai} \& {Tlatov}}]{Chatzistergos2020}
\bibinfo{author}{{Chatzistergos}, T.}, \bibinfo{author}{{Ermolli}, I.},
  \bibinfo{author}{{Krivova}, N.~A.}, \bibinfo{author}{{Solanki}, S.~K.},
  \bibinfo{author}{{Banerjee}, D.}, \bibinfo{author}{{Barata}, T.},
  \bibinfo{author}{{Belik}, M.}, \bibinfo{author}{{Gafeira}, R.},
  \bibinfo{author}{{Garcia}, A.}, \bibinfo{author}{{Hanaoka}, Y.},
  \bibinfo{author}{{Hegde}, M.}, \bibinfo{author}{{Klime{\v{s}}}, J.},
  \bibinfo{author}{{Korokhin}, V.~V.}, \bibinfo{author}{{Louren{\c{c}}o}, A.},
  \bibinfo{author}{{Malherbe}, J.-M.}, \bibinfo{author}{{Marchenko}, G.~P.},
  \bibinfo{author}{{Peixinho}, N.}, \bibinfo{author}{{Sakurai}, T.}, \&
  \bibinfo{author}{{Tlatov}, A.~G.} (\bibinfo{year}{2020}).
\newblock \bibinfo{title}{{Analysis of full-disc Ca II K spectroheliograms.
  III. Plage area composite series covering 1892-2019}}.
\newblock {\it \bibinfo{journal}{Astronomy \& Astrophysics}\/},  {\it
  \bibinfo{volume}{639}\/}, \bibinfo{pages}{A88}.
  \DOIprefix\doi{10.1051/0004-6361/202037746}.
  \href{http://arxiv.org/abs/2005.01435}{\tt arXiv:2005.01435}.
\bibitem[{{Chatzistergos} et~al.(2019){Chatzistergos}, {Ermolli}, {Solanki},
  {Krivova}, {Giorgi} \& {Yeo}}]{Chatzistergos2019}
\bibinfo{author}{{Chatzistergos}, T.}, \bibinfo{author}{{Ermolli}, I.},
  \bibinfo{author}{{Solanki}, S.~K.}, \bibinfo{author}{{Krivova}, N.~A.},
  \bibinfo{author}{{Giorgi}, F.}, \& \bibinfo{author}{{Yeo}, K.~L.}
  (\bibinfo{year}{2019}).
\newblock \bibinfo{title}{{Recovering the unsigned photospheric magnetic field
  from Ca II K observations}}.
\newblock {\it \bibinfo{journal}{Astronomy \& Astrophysics}\/},  {\it
  \bibinfo{volume}{626}\/}, \bibinfo{pages}{A114}.
  \DOIprefix\doi{10.1051/0004-6361/201935131}.
  \href{http://arxiv.org/abs/1905.03453}{\tt arXiv:1905.03453}.
\bibitem[{{Cliver} et~al.(1996){Cliver}, {Boriakoff} \& {Bounar}}]{Cliver1996}
\bibinfo{author}{{Cliver}, E.~W.}, \bibinfo{author}{{Boriakoff}, V.}, \&
  \bibinfo{author}{{Bounar}, K.~H.} (\bibinfo{year}{1996}).
\newblock \bibinfo{title}{{The 22-year cycle of geomagnetic and solar wind
  activity}}.
\newblock {\it \bibinfo{journal}{J. Geophys. Res.}\/},  {\it
  \bibinfo{volume}{101}\/}\bibinfo{issue}{(A12)},
  \bibinfo{pages}{27091--27110}. \DOIprefix\doi{10.1029/96JA02037}.
\bibitem[{{Crooker} et~al.(1977){Crooker}, {Feynman} \&
  {Gosling}}]{Crooker1977}
\bibinfo{author}{{Crooker}, N.~U.}, \bibinfo{author}{{Feynman}, J.}, \&
  \bibinfo{author}{{Gosling}, J.~T.} (\bibinfo{year}{1977}).
\newblock \bibinfo{title}{{On the high correlation between long-term averages
  of solar wind speed and geomagnetic activity}}.
\newblock {\it \bibinfo{journal}{J. Geophys. Res.}\/},  {\it
  \bibinfo{volume}{82}\/}\bibinfo{issue}{(13)}, \bibinfo{pages}{1933--1937}.
  \DOIprefix\doi{10.1029/JA082i013p01933}.
\bibitem[{{Dmitriev} et~al.(2013){Dmitriev}, {Suvorova} \&
  {Veselovsky}}]{Dmitriev2009}
\bibinfo{author}{{Dmitriev}, A.~V.}, \bibinfo{author}{{Suvorova}, A.~V.}, \&
  \bibinfo{author}{{Veselovsky}, I.~S.} (\bibinfo{year}{2013}).
\newblock \bibinfo{title}{{Statistical Characteristics of the Heliospheric
  Plasma and Magnetic Field at the Earth's Orbit during Four Solar Cycles
  20-23}}.
\newblock {\it \bibinfo{journal}{arXiv e-prints}\/},  (p.
  \bibinfo{pages}{arXiv:1301.2929}). \href{http://arxiv.org/abs/1301.2929}{\tt
  arXiv:1301.2929}.
\bibitem[{{Du}(2011)}]{Du2011}
\bibinfo{author}{{Du}, Z.~L.} (\bibinfo{year}{2011}).
\newblock \bibinfo{title}{{The correlation between solar and geomagnetic
  activity - Part 2: Long-term trends}}.
\newblock {\it \bibinfo{journal}{Annales Geophysicae}\/},  {\it
  \bibinfo{volume}{29}\/}\bibinfo{issue}{(8)}, \bibinfo{pages}{1341--1348}.
  \DOIprefix\doi{10.5194/angeo-29-1341-2011}.
\bibitem[{{Echer} et~al.(2004){Echer}, {Gonzalez}, {Gonzalez}, {Prestes},
  {Vieira}, {dal Lago}, {Guarnieri} \& {Schuch}}]{Echer2004}
\bibinfo{author}{{Echer}, E.}, \bibinfo{author}{{Gonzalez}, W.~D.},
  \bibinfo{author}{{Gonzalez}, A.~L.~C.}, \bibinfo{author}{{Prestes}, A.},
  \bibinfo{author}{{Vieira}, L.~E.~A.}, \bibinfo{author}{{dal Lago}, A.},
  \bibinfo{author}{{Guarnieri}, F.~L.}, \& \bibinfo{author}{{Schuch}, N.~J.}
  (\bibinfo{year}{2004}).
\newblock \bibinfo{title}{{Long-term correlation between solar and geomagnetic
  activity}}.
\newblock {\it \bibinfo{journal}{Journal of Atmospheric and Solar-Terrestrial
  Physics}\/},  {\it \bibinfo{volume}{66}\/}\bibinfo{issue}{(12)},
  \bibinfo{pages}{1019--1025}. \DOIprefix\doi{10.1016/j.jastp.2004.03.011}.
\bibitem[{{El-Borie}(2002)}]{ElBorie02}
\bibinfo{author}{{El-Borie}, M.~A.} (\bibinfo{year}{2002}).
\newblock \bibinfo{title}{{On Long-Term Periodicities In The Solar-Wind Ion
  Density and Speed Measurements During The Period 1973-2000}}.
\newblock {\it \bibinfo{journal}{Sol. Phys.}\/},  {\it
  \bibinfo{volume}{208}\/}\bibinfo{issue}{(2)}, \bibinfo{pages}{345--358}.
  \DOIprefix\doi{10.1023/A:1020585822820}.
\bibitem[{{Feldman} et~al.(1978){Feldman}, {Asbridge}, {Bame} \&
  {Gosling}}]{Feldman1978}
\bibinfo{author}{{Feldman}, W.~C.}, \bibinfo{author}{{Asbridge}, J.~R.},
  \bibinfo{author}{{Bame}, S.~J.}, \& \bibinfo{author}{{Gosling}, J.~T.}
  (\bibinfo{year}{1978}).
\newblock \bibinfo{title}{{Long-term variations of selected solar wind
  properties: Imp 6, 7, and 8 results}}.
\newblock {\it \bibinfo{journal}{J. Geophys. Res.}\/},  {\it
  \bibinfo{volume}{83}\/}\bibinfo{issue}{(A5)}, \bibinfo{pages}{2177--2189}.
  \DOIprefix\doi{10.1029/JA083iA05p02177}.
\bibitem[{{Feynman}(1982)}]{Feynman82}
\bibinfo{author}{{Feynman}, J.} (\bibinfo{year}{1982}).
\newblock \bibinfo{title}{{Feynman: Geomagnetic and solar wind cycles,
  1900-1975}}.
\newblock {\it \bibinfo{journal}{J. Geophys. Res.}\/},  {\it
  \bibinfo{volume}{87}\/}\bibinfo{issue}{(A8)}, \bibinfo{pages}{6153--6162}.
  \DOIprefix\doi{10.1029/JA087iA08p06153}.
\bibitem[{{Ghil} \& {Lucarini}(2020)}]{Ghil20}
\bibinfo{author}{{Ghil}, M.}, \& \bibinfo{author}{{Lucarini}, V.}
  (\bibinfo{year}{2020}).
\newblock \bibinfo{title}{{The physics of climate variability and climate
  change}}.
\newblock {\it \bibinfo{journal}{Reviews of Modern Physics}\/},  {\it
  \bibinfo{volume}{92}\/}\bibinfo{issue}{(3)}, \bibinfo{pages}{035002}.
  \DOIprefix\doi{10.1103/RevModPhys.92.035002}.
  \href{http://arxiv.org/abs/1910.00583}{\tt arXiv:1910.00583}.
\bibitem[{{Gonzalez} et~al.(1990){Gonzalez}, {Gonzalez} \&
  {Tsurutani}}]{Gonzalez1990}
\bibinfo{author}{{Gonzalez}, W.~D.}, \bibinfo{author}{{Gonzalez}, A.~L.~C.}, \&
  \bibinfo{author}{{Tsurutani}, B.~T.} (\bibinfo{year}{1990}).
\newblock \bibinfo{title}{{Dual-peak solar cycle distribution of intense
  geomagnetic storms}}.
\newblock {\it \bibinfo{journal}{Planetary and Space Science}\/},  {\it
  \bibinfo{volume}{38}\/}\bibinfo{issue}{(2)}, \bibinfo{pages}{181--187}.
  \DOIprefix\doi{10.1016/0032-0633(90)90082-2}.
\bibitem[{{Hathaway}(2010)}]{Hathaway2010}
\bibinfo{author}{{Hathaway}, D.~H.} (\bibinfo{year}{2010}).
\newblock \bibinfo{title}{{The Solar Cycle}}.
\newblock {\it \bibinfo{journal}{Living Reviews in Solar Physics}\/},  {\it
  \bibinfo{volume}{7}\/}\bibinfo{issue}{(1)}, \bibinfo{pages}{1}.
  \DOIprefix\doi{10.12942/lrsp-2010-1}.
\bibitem[{{Hirshberg}(1973)}]{Hirshberg73}
\bibinfo{author}{{Hirshberg}, J.} (\bibinfo{year}{1973}).
\newblock \bibinfo{title}{{The Solar Wind Cycle, the Sunspot Cycle, and the
  Corona}}.
\newblock {\it \bibinfo{journal}{Astrophys. Space Sci.}\/},  {\it
  \bibinfo{volume}{20}\/}\bibinfo{issue}{(2)}, \bibinfo{pages}{473--481}.
  \DOIprefix\doi{10.1007/BF00642216}.
\bibitem[{{Intriligator}(1974)}]{Intriligator74}
\bibinfo{author}{{Intriligator}, D.~S.} (\bibinfo{year}{1974}).
\newblock \bibinfo{title}{{Evidence of Solar-Cycle Variations in the Solar
  Wind}}.
\newblock {\it \bibinfo{journal}{The Astrophysical Journal Letters}\/},  {\it
  \bibinfo{volume}{188}\/}, \bibinfo{pages}{L23--L26}.
  \DOIprefix\doi{10.1086/181422}.
\bibitem[{{Kane}(2003)}]{Kane2003}
\bibinfo{author}{{Kane}, R.~P.} (\bibinfo{year}{2003}).
\newblock \bibinfo{title}{{Lags, hysteresis, and double peaks between cosmic
  rays and solar activity}}.
\newblock {\it \bibinfo{journal}{Journal of Geophysical Research (Space
  Physics)}\/},  {\it \bibinfo{volume}{108}\/}\bibinfo{issue}{(A10)},
  \bibinfo{pages}{1379}. \DOIprefix\doi{10.1029/2003JA009995}.
\bibitem[{{Katsavrias} et~al.(2012){Katsavrias}, {Preka-Papadema} \&
  {Moussas}}]{Katsavrias12}
\bibinfo{author}{{Katsavrias}, C.}, \bibinfo{author}{{Preka-Papadema}, P.}, \&
  \bibinfo{author}{{Moussas}, X.} (\bibinfo{year}{2012}).
\newblock \bibinfo{title}{{Wavelet Analysis on Solar Wind Parameters and
  Geomagnetic Indices}}.
\newblock {\it \bibinfo{journal}{Sol. Phys.}\/},  {\it
  \bibinfo{volume}{280}\/}\bibinfo{issue}{(2)}, \bibinfo{pages}{623--640}.
  \DOIprefix\doi{10.1007/s11207-012-0078-6}.
  \href{http://arxiv.org/abs/1205.2229}{\tt arXiv:1205.2229}.
\bibitem[{{King}(1979)}]{King1979}
\bibinfo{author}{{King}, J.~H.} (\bibinfo{year}{1979}).
\newblock \bibinfo{title}{{Solar cycle variations in IMF intensity}}.
\newblock {\it \bibinfo{journal}{J. Geophys. Res.}\/},  {\it
  \bibinfo{volume}{84}\/}\bibinfo{issue}{(A10)}, \bibinfo{pages}{5938--5940}.
  \DOIprefix\doi{10.1029/JA084iA10p05938}.
\bibitem[{{King} \& {Papitashvili}(2005)}]{King2005}
\bibinfo{author}{{King}, J.~H.}, \& \bibinfo{author}{{Papitashvili}, N.~E.}
  (\bibinfo{year}{2005}).
\newblock \bibinfo{title}{{Solar wind spatial scales in and comparisons of
  hourly Wind and ACE plasma and magnetic field data}}.
\newblock {\it \bibinfo{journal}{Journal of Geophysical Research (Space
  Physics)}\/},  {\it \bibinfo{volume}{110}\/}\bibinfo{issue}{(A2)},
  \bibinfo{pages}{A02104}. \DOIprefix\doi{10.1029/2004JA010649}.
\bibitem[{{K{\"o}hnlein}(1996)}]{Kohnlein96}
\bibinfo{author}{{K{\"o}hnlein}, W.} (\bibinfo{year}{1996}).
\newblock \bibinfo{title}{{Cross-correlation of solar wind parameters with
  sunspots (`Long-term variations') at 1 AU during cycles 21 and 22}}.
\newblock {\it \bibinfo{journal}{Astrophys. Space Sci.}\/},  {\it
  \bibinfo{volume}{245}\/}\bibinfo{issue}{(1)}, \bibinfo{pages}{81--88}.
  \DOIprefix\doi{10.1007/BF00637804}.
\bibitem[{{Koldobskiy} et~al.(2022){Koldobskiy}, {Kähkönen}, {Hofer},
  {Krivova}, {Kovaltsov} \& {Usoskin}}]{Koldobskiy2022}
\bibinfo{author}{{Koldobskiy}, S.~A.}, \bibinfo{author}{{Kähkönen}, R.},
  \bibinfo{author}{{Hofer}, B.}, \bibinfo{author}{{Krivova}, N.~A.},
  \bibinfo{author}{{Kovaltsov}, G.~A.}, \& \bibinfo{author}{{Usoskin}, I.~G.}
  (\bibinfo{year}{2022}).
\newblock \bibinfo{title}{{Time Lag Between Cosmic-Ray and Solar Variability:
  Sunspot Numbers and Open Solar Magnetic Flux}}.
\newblock {\it \bibinfo{journal}{Sol. Phys.}\/},  {\it
  \bibinfo{volume}{297}\/}\bibinfo{issue}{(38)}.
  \DOIprefix\doi{10.1007/s11207-022-01970-1}.
\bibitem[{{Laperre} et~al.(2020){Laperre}, {Amaya} \& {Lapenta}}]{Laperre2020}
\bibinfo{author}{{Laperre}, B.}, \bibinfo{author}{{Amaya}, J.}, \&
  \bibinfo{author}{{Lapenta}, G.} (\bibinfo{year}{2020}).
\newblock \bibinfo{title}{{Dynamic Time Warping as a New Evaluation for Dst
  Forecast with Machine Learning}}.
\newblock {\it \bibinfo{journal}{Frontiers in Astronomy and Space Sciences}\/},
   {\it \bibinfo{volume}{7}\/}, \bibinfo{pages}{39}.
  \DOIprefix\doi{10.3389/fspas.2020.00039}.
  \href{http://arxiv.org/abs/2006.04667}{\tt arXiv:2006.04667}.
\bibitem[{{Lepping} et~al.(1995){Lepping}, {Ac{\~{u}}na}, {Burlaga}, {Farrell},
  {Slavin}, {Schatten}, {Mariani}, {Ness}, {Neubauer}, {Whang}, {Byrnes},
  {Kennon}, {Panetta}, {Scheifele} \& {Worley}}]{Lepping95}
\bibinfo{author}{{Lepping}, R.~P.}, \bibinfo{author}{{Ac{\~{u}}na}, M.~H.},
  \bibinfo{author}{{Burlaga}, L.~F.}, \bibinfo{author}{{Farrell}, W.~M.},
  \bibinfo{author}{{Slavin}, J.~A.}, \bibinfo{author}{{Schatten}, K.~H.},
  \bibinfo{author}{{Mariani}, F.}, \bibinfo{author}{{Ness}, N.~F.},
  \bibinfo{author}{{Neubauer}, F.~M.}, \bibinfo{author}{{Whang}, Y.~C.},
  \bibinfo{author}{{Byrnes}, J.~B.}, \bibinfo{author}{{Kennon}, R.~S.},
  \bibinfo{author}{{Panetta}, P.~V.}, \bibinfo{author}{{Scheifele}, J.}, \&
  \bibinfo{author}{{Worley}, E.~M.} (\bibinfo{year}{1995}).
\newblock \bibinfo{title}{{The Wind Magnetic Field Investigation}}.
\newblock {\it \bibinfo{journal}{Space Science Reviews}\/},  {\it
  \bibinfo{volume}{71}\/}\bibinfo{issue}{(1-4)}, \bibinfo{pages}{207--229}.
  \DOIprefix\doi{10.1007/BF00751330}.
\bibitem[{{Li} et~al.(2017){Li}, {Zhang} \& {Feng}}]{Li17}
\bibinfo{author}{{Li}, K.~J.}, \bibinfo{author}{{Zhang}, J.}, \&
  \bibinfo{author}{{Feng}, W.} (\bibinfo{year}{2017}).
\newblock \bibinfo{title}{{Periodicity for 50 yr of daily solar wind
  velocity}}.
\newblock {\it \bibinfo{journal}{Mon. Not. R. Astron. Soc.}\/},  {\it
  \bibinfo{volume}{472}\/}\bibinfo{issue}{(1)}, \bibinfo{pages}{289--294}.
  \DOIprefix\doi{10.1093/mnras/stx1904}.
\bibitem[{{Li} et~al.(2016){Li}, {Zhanng} \& {Feng}}]{Li2016}
\bibinfo{author}{{Li}, K.~J.}, \bibinfo{author}{{Zhanng}, J.}, \&
  \bibinfo{author}{{Feng}, W.} (\bibinfo{year}{2016}).
\newblock \bibinfo{title}{{A Statistical Analysis of 50 Years of Daily Solar
  Wind Velocity Data}}.
\newblock {\it \bibinfo{journal}{The Astronomical Journal}\/},  {\it
  \bibinfo{volume}{151}\/}\bibinfo{issue}{(5)}, \bibinfo{pages}{128}.
  \DOIprefix\doi{10.3847/0004-6256/151/5/128}.
\bibitem[{{Lomb}(1976)}]{Lomb76}
\bibinfo{author}{{Lomb}, N.~R.} (\bibinfo{year}{1976}).
\newblock \bibinfo{title}{{Least-Squares Frequency Analysis of Unequally Spaced
  Data}}.
\newblock {\it \bibinfo{journal}{Astrophys. Space Sci.}\/},  {\it
  \bibinfo{volume}{39}\/}\bibinfo{issue}{(2)}, \bibinfo{pages}{447--462}.
  \DOIprefix\doi{10.1007/BF00648343}.
\bibitem[{{Neugebauer}(1981)}]{Neugebauer1981}
\bibinfo{author}{{Neugebauer}, M.} (\bibinfo{year}{1981}).
\newblock \bibinfo{title}{{Observations of Solar-Wind Helium}}.
\newblock {\it \bibinfo{journal}{Fundamentals of Cosmic Physics}\/},  {\it
  \bibinfo{volume}{7}\/}, \bibinfo{pages}{131--199}.
\bibitem[{{North} et~al.(1981){North}, {Cahalan} \& {Coakley}}]{North81}
\bibinfo{author}{{North}, G.~R.}, \bibinfo{author}{{Cahalan}, R.~F.}, \&
  \bibinfo{author}{{Coakley}, J., James~A.} (\bibinfo{year}{1981}).
\newblock \bibinfo{title}{{Energy Balance Climate Models (Paper 80R1502)}}.
\newblock {\it \bibinfo{journal}{Reviews of Geophysics and Space Physics}\/},
  {\it \bibinfo{volume}{19}\/}, \bibinfo{pages}{91--121}.
  \DOIprefix\doi{10.1029/RG019i001p00091}.
\bibitem[{{Ogilvie} \& {Hirshberg}(1974)}]{Ogilvie1974}
\bibinfo{author}{{Ogilvie}, K.~W.}, \& \bibinfo{author}{{Hirshberg}, J.}
  (\bibinfo{year}{1974}).
\newblock \bibinfo{title}{{The solar cycle variation of the solar wind helium
  abundance}}.
\newblock {\it \bibinfo{journal}{J. Geophys. Res.}\/},  {\it
  \bibinfo{volume}{79}\/}\bibinfo{issue}{(31)}, \bibinfo{pages}{4595--4602}.
  \DOIprefix\doi{10.1029/JA079i031p04595}.
\bibitem[{{Ogilvie} et~al.(1977){Ogilvie}, {von Rosenvinge} \&
  {Durney}}]{Ogilvie77}
\bibinfo{author}{{Ogilvie}, K.~W.}, \bibinfo{author}{{von Rosenvinge}, T.}, \&
  \bibinfo{author}{{Durney}, A.~C.} (\bibinfo{year}{1977}).
\newblock \bibinfo{title}{{International Sun-Earth Explorer: A Three-Spacecraft
  Program}}.
\newblock {\it \bibinfo{journal}{Science}\/},  {\it
  \bibinfo{volume}{198}\/}\bibinfo{issue}{(4313)}, \bibinfo{pages}{131--138}.
  \DOIprefix\doi{10.1126/science.198.4313.131}.
\bibitem[{{{\"O}zg{\"u}{\c{c}}} et~al.(2016){{\"O}zg{\"u}{\c{c}}}, {Kilcik},
  {Georgieva} \& {Kirov}}]{Ozguc2016}
\bibinfo{author}{{{\"O}zg{\"u}{\c{c}}}, A.}, \bibinfo{author}{{Kilcik}, A.},
  \bibinfo{author}{{Georgieva}, K.}, \& \bibinfo{author}{{Kirov}, B.}
  (\bibinfo{year}{2016}).
\newblock \bibinfo{title}{{Temporal Offsets Between Maximum CME Speed Index and
  Solar, Geomagnetic, and Interplanetary Indicators During Solar Cycle 23 and
  the Ascending Phase of Cycle 24}}.
\newblock {\it \bibinfo{journal}{Solar Physics}\/},  {\it
  \bibinfo{volume}{291}\/}\bibinfo{issue}{(5)}, \bibinfo{pages}{1533--1546}.
  \DOIprefix\doi{10.1007/s11207-016-0909-y}.
  \href{http://arxiv.org/abs/1604.05941}{\tt arXiv:1604.05941}.
\bibitem[{{{\"O}zg{\"u}{\c{c}}} et~al.(2012){{\"O}zg{\"u}{\c{c}}}, {Kilcik} \&
  {Rozelot}}]{Ozguc2012}
\bibinfo{author}{{{\"O}zg{\"u}{\c{c}}}, A.}, \bibinfo{author}{{Kilcik}, A.}, \&
  \bibinfo{author}{{Rozelot}, J.} (\bibinfo{year}{2012}).
\newblock \bibinfo{title}{Effects of hysteresis between maximum cme speed index
  and typical solar activity indicators during cycle 23}.
\newblock {\it \bibinfo{journal}{Solar Physics}\/},  {\it
  \bibinfo{volume}{281}\/}, \bibinfo{pages}{839--846}.
  \DOIprefix\doi{10.1007/s11207-012-0087-5}.
\bibitem[{{Parks}(2004)}]{Parks04}
\bibinfo{author}{{Parks}, G.~K.} (\bibinfo{year}{2004}).
\newblock {\it \bibinfo{title}{{Physics of space plasmas : an
  introduction}}\/}.
\newblock \bibinfo{publisher}{Westview Press, Boulder Colorado}.
\bibitem[{{Perri} et~al.(2021){Perri}, {Brun}, {Strugarek} \&
  {R{\'e}ville}}]{Perri2021}
\bibinfo{author}{{Perri}, B.}, \bibinfo{author}{{Brun}, A.~S.},
  \bibinfo{author}{{Strugarek}, A.}, \& \bibinfo{author}{{R{\'e}ville}, V.}
  (\bibinfo{year}{2021}).
\newblock \bibinfo{title}{{Dynamical Coupling of a Mean-field Dynamo and Its
  Wind: Feedback Loop over a Stellar Activity Cycle}}.
\newblock {\it \bibinfo{journal}{The Astrophysical Journal}\/},  {\it
  \bibinfo{volume}{910}\/}\bibinfo{issue}{(1)}, \bibinfo{pages}{50}.
  \DOIprefix\doi{10.3847/1538-4357/abe2ac}.
  \href{http://arxiv.org/abs/2102.01416}{\tt arXiv:2102.01416}.
\bibitem[{{Prabhakaran Nayar} et~al.(2002){Prabhakaran Nayar}, {Radhika},
  {Revathy} \& {Ramadas}}]{Prabhakaran2002}
\bibinfo{author}{{Prabhakaran Nayar}, S.~R.}, \bibinfo{author}{{Radhika},
  V.~N.}, \bibinfo{author}{{Revathy}, K.}, \& \bibinfo{author}{{Ramadas}, V.}
  (\bibinfo{year}{2002}).
\newblock \bibinfo{title}{{Wavelet Analysis of solar, solar wind and
  geomagnetic parameters}}.
\newblock {\it \bibinfo{journal}{Sol. Phys.}\/},  {\it
  \bibinfo{volume}{208}\/}\bibinfo{issue}{(2)}, \bibinfo{pages}{359--373}.
  \DOIprefix\doi{10.1023/A:1020565831926}.
\bibitem[{{Reda} et~al.(2022){Reda}, {Giovannelli}, {Alberti}, {Berrilli},
  {Bertello}, {Del Moro}, {Di Mauro}, {Giobbi} \& {Penza}}]{Reda2022}
\bibinfo{author}{{Reda}, R.}, \bibinfo{author}{{Giovannelli}, L.},
  \bibinfo{author}{{Alberti}, T.}, \bibinfo{author}{{Berrilli}, F.},
  \bibinfo{author}{{Bertello}, L.}, \bibinfo{author}{{Del Moro}, D.},
  \bibinfo{author}{{Di Mauro}, M.~P.}, \bibinfo{author}{{Giobbi}, P.}, \&
  \bibinfo{author}{{Penza}, V.} (\bibinfo{year}{2022}).
\newblock \bibinfo{title}{{The exoplanetary magnetosphere extension in Sun-like
  stars based on the solar wind - solar UV relation}}.
\newblock {\it \bibinfo{journal}{arXiv e-prints}\/},  (p.
  \bibinfo{pages}{arXiv:2203.01554}).
  \href{http://arxiv.org/abs/2203.01554}{\tt arXiv:2203.01554}.
\bibitem[{{Richardson} \& {Cane}(2012)}]{Richardson2012}
\bibinfo{author}{{Richardson}, I.~G.}, \& \bibinfo{author}{{Cane}, H.~V.}
  (\bibinfo{year}{2012}).
\newblock \bibinfo{title}{{Near-earth solar wind flows and related geomagnetic
  activity during more than four solar cycles (1963-2011)}}.
\newblock {\it \bibinfo{journal}{Journal of Space Weather and Space
  Climate}\/},  {\it \bibinfo{volume}{2}\/}, \bibinfo{pages}{A02}.
  \DOIprefix\doi{10.1051/swsc/2012003}.
\bibitem[{{Richardson} et~al.(2000){Richardson}, {Cliver} \&
  {Cane}}]{Richardson2000}
\bibinfo{author}{{Richardson}, I.~G.}, \bibinfo{author}{{Cliver}, E.~W.}, \&
  \bibinfo{author}{{Cane}, H.~V.} (\bibinfo{year}{2000}).
\newblock \bibinfo{title}{{Sources of geomagnetic activity over the solar
  cycle: Relative importance of coronal mass ejections, high-speed streams, and
  slow solar wind}}.
\newblock {\it \bibinfo{journal}{J. Geophys. Res.}\/},  {\it
  \bibinfo{volume}{105}\/}\bibinfo{issue}{(A8)},
  \bibinfo{pages}{18,203--18,213}. \DOIprefix\doi{10.1029/1999JA000400}.
\bibitem[{{Ross} \& {Chaplin}(2019)}]{Ross2019}
\bibinfo{author}{{Ross}, E.}, \& \bibinfo{author}{{Chaplin}, W.~J.}
  (\bibinfo{year}{2019}).
\newblock \bibinfo{title}{{The Behaviour of Galactic Cosmic-Ray Intensity
  During Solar Activity Cycle 24}}.
\newblock {\it \bibinfo{journal}{Solar Physics}\/},  {\it
  \bibinfo{volume}{294}\/}\bibinfo{issue}{(1)}, \bibinfo{pages}{8}.
  \DOIprefix\doi{10.1007/s11207-019-1397-7}.
\bibitem[{{Samara} et~al.(2022){Samara}, {Laperre}, {Kieokaew}, {Temmer},
  {Verbeke}, {Rodriguez}, {Magdaleni{\'c}} \& {Poedts}}]{Samara2022}
\bibinfo{author}{{Samara}, E.}, \bibinfo{author}{{Laperre}, B.},
  \bibinfo{author}{{Kieokaew}, R.}, \bibinfo{author}{{Temmer}, M.},
  \bibinfo{author}{{Verbeke}, C.}, \bibinfo{author}{{Rodriguez}, L.},
  \bibinfo{author}{{Magdaleni{\'c}}, J.}, \& \bibinfo{author}{{Poedts}, S.}
  (\bibinfo{year}{2022}).
\newblock \bibinfo{title}{{Dynamic Time Warping as a Means of Assessing Solar
  Wind Time Series}}.
\newblock {\it \bibinfo{journal}{The Astrophysical Journal}\/},  {\it
  \bibinfo{volume}{927}\/}\bibinfo{issue}{(2)}, \bibinfo{pages}{187}.
  \DOIprefix\doi{10.3847/1538-4357/ac4af6}.
  \href{http://arxiv.org/abs/2109.07873}{\tt arXiv:2109.07873}.
\bibitem[{{Samsonov} et~al.(2019){Samsonov}, {Bogdanova}, {Branduardi-Raymont},
  {Safrankova}, {Nemecek} \& {Park}}]{Samsonov2019}
\bibinfo{author}{{Samsonov}, A.~A.}, \bibinfo{author}{{Bogdanova}, Y.~V.},
  \bibinfo{author}{{Branduardi-Raymont}, G.}, \bibinfo{author}{{Safrankova},
  J.}, \bibinfo{author}{{Nemecek}, Z.}, \& \bibinfo{author}{{Park}, J.~S.}
  (\bibinfo{year}{2019}).
\newblock \bibinfo{title}{{Long-Term Variations in Solar Wind Parameters,
  Magnetopause Location, and Geomagnetic Activity Over the Last Five Solar
  Cycles}}.
\newblock {\it \bibinfo{journal}{Journal of Geophysical Research (Space
  Physics)}\/},  {\it \bibinfo{volume}{124}\/}\bibinfo{issue}{(6)},
  \bibinfo{pages}{4049--4063}. \DOIprefix\doi{10.1029/2018JA026355}.
\bibitem[{{Sarp} et~al.(2019){Sarp}, {Kilcik}, {Yurchyshyn}, {Ozguc} \&
  {Rozelot}}]{Sarp2019}
\bibinfo{author}{{Sarp}, V.}, \bibinfo{author}{{Kilcik}, A.},
  \bibinfo{author}{{Yurchyshyn}, V.}, \bibinfo{author}{{Ozguc}, A.}, \&
  \bibinfo{author}{{Rozelot}, J.-P.} (\bibinfo{year}{2019}).
\newblock \bibinfo{title}{{Cosmic Ray Modulation with the Maximum CME Speed
  Index During Solar Cycles 23 and 24}}.
\newblock {\it \bibinfo{journal}{Solar Physics}\/},  {\it
  \bibinfo{volume}{294}\/}\bibinfo{issue}{(7)}, \bibinfo{pages}{86}.
  \DOIprefix\doi{10.1007/s11207-019-1481-z}.
\bibitem[{{Scargle}(1982)}]{Scargle82}
\bibinfo{author}{{Scargle}, J.~D.} (\bibinfo{year}{1982}).
\newblock \bibinfo{title}{{Studies in astronomical time series analysis. II.
  Statistical aspects of spectral analysis of unevenly spaced data.}}
\newblock {\it \bibinfo{journal}{Astrophys J.}\/},  {\it
  \bibinfo{volume}{263}\/}, \bibinfo{pages}{835--853}.
  \DOIprefix\doi{10.1086/160554}.
\bibitem[{{Schrijver} et~al.(1989){Schrijver}, {Cote}, {Zwaan} \&
  {Saar}}]{Schrijver1989}
\bibinfo{author}{{Schrijver}, C.~J.}, \bibinfo{author}{{Cote}, J.},
  \bibinfo{author}{{Zwaan}, C.}, \& \bibinfo{author}{{Saar}, S.~H.}
  (\bibinfo{year}{1989}).
\newblock \bibinfo{title}{{Relations between the Photospheric Magnetic Field
  and the Emission from the Outer Atmospheres of Cool Stars. I. The Solar CA II
  K Line Core Emission}}.
\newblock {\it \bibinfo{journal}{The Astrophysical Journal}\/},  {\it
  \bibinfo{volume}{337}\/}, \bibinfo{pages}{964--976}.
  \DOIprefix\doi{10.1086/167168}.
\bibitem[{{Schwabe}(1844)}]{Schwabe1844}
\bibinfo{author}{{Schwabe}, H.} (\bibinfo{year}{1844}).
\newblock \bibinfo{title}{{Solar Observations During 1843}}.
\newblock {\it \bibinfo{journal}{Astronomische Nachrichten}\/},  {\it
  \bibinfo{volume}{21}\/}\bibinfo{issue}{(495)}, \bibinfo{pages}{233--236}.
  \DOIprefix\doi{10.1002/asna.18440211505}.
\bibitem[{{Siscoe} et~al.(1978){Siscoe}, {Crooker} \&
  {Christopher}}]{Siscoe1978}
\bibinfo{author}{{Siscoe}, G.~L.}, \bibinfo{author}{{Crooker}, N.~U.}, \&
  \bibinfo{author}{{Christopher}, L.} (\bibinfo{year}{1978}).
\newblock \bibinfo{title}{{A solar cycle variation of the interplanetary
  magnetic field.}}
\newblock {\it \bibinfo{journal}{Sol. Phys.}\/},  {\it
  \bibinfo{volume}{56}\/}\bibinfo{issue}{(2)}, \bibinfo{pages}{449--461}.
  \DOIprefix\doi{10.1007/BF00152484}.
\bibitem[{{Stone} et~al.(1990){Stone}, {Burlaga}, {Cummings}, {Feldman},
  {Frain}, {Geiss}, {Gloeckler}, {Gold}, {Hovestadt}, {Krimigis}, {Mason},
  {McComas}, {Mewaldt}, {Simpson}, {von Rosenvinge} \& {Wiedenbeck}}]{Stone90}
\bibinfo{author}{{Stone}, E.~C.}, \bibinfo{author}{{Burlaga}, L.~F.},
  \bibinfo{author}{{Cummings}, A.~C.}, \bibinfo{author}{{Feldman}, W.~C.},
  \bibinfo{author}{{Frain}, W.~E.}, \bibinfo{author}{{Geiss}, J.},
  \bibinfo{author}{{Gloeckler}, G.}, \bibinfo{author}{{Gold}, R.~E.},
  \bibinfo{author}{{Hovestadt}, D.}, \bibinfo{author}{{Krimigis}, S.~M.},
  \bibinfo{author}{{Mason}, G.~M.}, \bibinfo{author}{{McComas}, D.},
  \bibinfo{author}{{Mewaldt}, R.~A.}, \bibinfo{author}{{Simpson}, J.~A.},
  \bibinfo{author}{{von Rosenvinge}, T.~T.}, \& \bibinfo{author}{{Wiedenbeck},
  M.~E.} (\bibinfo{year}{1990}).
\newblock \bibinfo{title}{{The advanced composition explorer}}.
\newblock In \bibinfo{editor}{W.~V. {Jones}}, \bibinfo{editor}{F.~J. {Kerr}},
  \& \bibinfo{editor}{J.~F. {Ormes}} (Eds.), {\it \bibinfo{booktitle}{Particle
  Astrophysics - The NASA Cosmic Ray Program for the 1990s and Beyond}\/} (pp.
  \bibinfo{pages}{48--57}).
\newblock volume \bibinfo{volume}{203} of {\it \bibinfo{series}{American
  Institute of Physics Conference Series}\/}.
\newblock \DOIprefix\doi{10.1063/1.39173}.
\bibitem[{{Tlatov} et~al.(2009){Tlatov}, {Pevtsov} \& {Singh}}]{Tlatov2009}
\bibinfo{author}{{Tlatov}, A.~G.}, \bibinfo{author}{{Pevtsov}, A.~A.}, \&
  \bibinfo{author}{{Singh}, J.} (\bibinfo{year}{2009}).
\newblock \bibinfo{title}{{A New Method of Calibration of Photographic Plates
  from Three Historic Data Sets}}.
\newblock {\it \bibinfo{journal}{Sol. Phys.}\/},  {\it
  \bibinfo{volume}{255}\/}\bibinfo{issue}{(2)}, \bibinfo{pages}{239--251}.
  \DOIprefix\doi{10.1007/s11207-009-9326-9}.
\bibitem[{{Tsurutani} et~al.(2006){Tsurutani}, {Gonzalez}, {Gonzalez},
  {Guarnieri}, {Gopalswamy}, {Grande}, {Kamide}, {Kasahara}, {Lu}, {Mann},
  {McPherron}, {Soraas} \& {Vasyliunas}}]{Tsurutani2006}
\bibinfo{author}{{Tsurutani}, B.~T.}, \bibinfo{author}{{Gonzalez}, W.~D.},
  \bibinfo{author}{{Gonzalez}, A. L.~C.}, \bibinfo{author}{{Guarnieri}, F.~L.},
  \bibinfo{author}{{Gopalswamy}, N.}, \bibinfo{author}{{Grande}, M.},
  \bibinfo{author}{{Kamide}, Y.}, \bibinfo{author}{{Kasahara}, Y.},
  \bibinfo{author}{{Lu}, G.}, \bibinfo{author}{{Mann}, I.},
  \bibinfo{author}{{McPherron}, R.}, \bibinfo{author}{{Soraas}, F.}, \&
  \bibinfo{author}{{Vasyliunas}, V.} (\bibinfo{year}{2006}).
\newblock \bibinfo{title}{{Corotating solar wind streams and recurrent
  geomagnetic activity: A review}}.
\newblock {\it \bibinfo{journal}{Journal of Geophysical Research (Space
  Physics)}\/},  {\it \bibinfo{volume}{111}\/}\bibinfo{issue}{(A7)},
  \bibinfo{pages}{A07S01}. \DOIprefix\doi{10.1029/2005JA011273}.
\bibitem[{{Usoskin}(2017)}]{Usoskin2017}
\bibinfo{author}{{Usoskin}, I.~G.} (\bibinfo{year}{2017}).
\newblock \bibinfo{title}{{A history of solar activity over millennia}}.
\newblock {\it \bibinfo{journal}{Living Reviews in Solar Physics}\/},  {\it
  \bibinfo{volume}{14}\/}\bibinfo{issue}{(1)}, \bibinfo{pages}{3}.
  \DOIprefix\doi{10.1007/s41116-017-0006-9}.
\bibitem[{{Vecchio} et~al.(2017){Vecchio}, {Lepreti}, {Laurenza}, {Alberti} \&
  {Carbone}}]{Vecchio17}
\bibinfo{author}{{Vecchio}, A.}, \bibinfo{author}{{Lepreti}, F.},
  \bibinfo{author}{{Laurenza}, M.}, \bibinfo{author}{{Alberti}, T.}, \&
  \bibinfo{author}{{Carbone}, V.} (\bibinfo{year}{2017}).
\newblock \bibinfo{title}{{Connection between solar activity cycles and grand
  minima generation}}.
\newblock {\it \bibinfo{journal}{Astronomy \& Astrophysics}\/},  {\it
  \bibinfo{volume}{599}\/}, \bibinfo{pages}{A58}.
  \DOIprefix\doi{10.1051/0004-6361/201629758}.
\bibitem[{{Venzmer} \& {Bothmer}(2018)}]{Venzmer2018}
\bibinfo{author}{{Venzmer}, M.~S.}, \& \bibinfo{author}{{Bothmer}, V.}
  (\bibinfo{year}{2018}).
\newblock \bibinfo{title}{{Solar-wind predictions for the Parker Solar Probe
  orbit. Near-Sun extrapolations derived from an empirical solar-wind model
  based on Helios and OMNI observations}}.
\newblock {\it \bibinfo{journal}{Astronomy \& Astrophysics}\/},  {\it
  \bibinfo{volume}{611}\/}, \bibinfo{pages}{A36}.
  \DOIprefix\doi{10.1051/0004-6361/201731831}.
  \href{http://arxiv.org/abs/1711.07534}{\tt arXiv:1711.07534}.

\end{thebibliography}

\end{document}